\documentstyle[emulateapj,apjfonts,epsf]{article}

\received{}
\accepted{}
\journalid{}{}
\articleid{}{}

\slugcomment{Accepted for publication by {\it The Astrophysical Journal Letters}
July 10, 2001}

\lefthead{Camilo et al.}
\righthead{PSR J1016$-$5857: a Young and Energetic Pulsar}

\begin{document}

%
%

\def\egret{{\sl EGRET\/}}
\def\einstein{{\sl Einstein}}

\def\psr{PSR~J1016$-$5857}
\def\snr{G284.3$-$1.8}
\def\egr{3EG~J1013$-$5915}

\title{\psr: a young radio pulsar with possible supernova remnant,
X-ray, and $\gamma$-ray associations}

\author{F.~Camilo,\altaffilmark{1} 
  J.~F.~Bell,\altaffilmark{2}
  R.~N.~Manchester,\altaffilmark{2}
  A.~G.~Lyne,\altaffilmark{3}
  A.~Possenti,\altaffilmark{4}
  M.~Kramer,\altaffilmark{3}
  V.~M.~Kaspi,\altaffilmark{5}
  I.~H.~Stairs,\altaffilmark{6}
  N.~D'Amico,\altaffilmark{4}
  G.~Hobbs,\altaffilmark{3}
  E.~V.~Gotthelf,\altaffilmark{1}
  and B.~M.~Gaensler\altaffilmark{7} }
\medskip
\altaffiltext{1}{Columbia Astrophysics Laboratory, Columbia University,
  550 West 120th Street, New York, NY~10027}
\altaffiltext{2}{Australia Telescope National Facility, CSIRO,
  P.O.~Box~76, Epping, NSW~1710, Australia}
\altaffiltext{3}{University of Manchester, Jodrell Bank Observatory,
  Macclesfield, Cheshire, SK11~9DL, UK}
\altaffiltext{4}{Osservatorio Astronomico di Bologna, via Ranzani 1,
  40127 Bologna, Italy}
\altaffiltext{5}{Physics Department, McGill University, 3600 University
  Street, Montreal, Quebec, H3A~2T8, Canada}
\altaffiltext{6}{National Radio Astronomy Observatory, P.O. Box 2,
  Green Bank, WV~24944}
\altaffiltext{7}{Center for Space Research, Massachusetts Institute of
  Technology, Cambridge, MA~02139}

\begin{abstract}
We report the discovery of a young and energetic pulsar in the Parkes
multibeam survey of the Galactic plane. \psr\ has a rotation period of
$107$\,ms and period derivative of $8.0\times10^{-14}$, implying a
characteristic age of 21\,kyr and spin-down luminosity of
$2.6\times10^{36}$\,erg\,s$^{-1}$.  The pulsar is located just outside,
and possibly interacting with, the shell supernova remnant \snr.
Archival X-ray data show a source near the pulsar position which is
consistent with emission from a pulsar wind nebula.  The pulsar is also
located inside the error box of the unidentified \egret\ source \egr,
for which it represents a plausible counterpart.

\end{abstract}


\section{Introduction}\label{sec:intro}

Observations of young rotation-powered pulsars offer unique
opportunities to illuminate several fundamental questions about the
physics of neutron stars and supernova remnants (SNRs).  In particular,
synchrotron emission from a pulsar wind nebula (PWN) can be used to
determine properties of the relativistic pulsar wind, to probe the
density of the ambient medium, and to study the evolution of SNR ejecta
and their interaction with the local ISM.  Confirmed associations
between pulsars and SNRs also provide the opportunity to directly
constrain the birth properties of neutron stars, including their space
velocities and initial rotation periods.  Such associations are often
difficult to judge on the basis of the distance and age estimates for
both objects, and a very powerful method of confirming an association
is the observation of interaction between pulsar/PWN and SNR.  However,
young pulsars are rare, and cases where such interaction is apparent
are rarer still.

One of the main aims of the Parkes multibeam survey of the Galactic
plane (\cite{mlc+01}) is precisely to detect young pulsars, and at
least $\sim30$ of the 600 pulsars discovered to date are youthful, with
characteristic age $\tau_c < 100$\,kyr.  Several of these are energetic
as well (e.g., \cite{dkm+01}).  In this Letter we report on one such
object, which archival radio, X-ray, and $\gamma$-ray observations
suggest is remarkable.

\section{Radio Pulsar J1016$-$5857}\label{sec:psr}

\psr\ was discovered in data acquired on 1999 March 15, collected in
the course of the pulsar multibeam survey (\cite{mlc+01}) at the 64-m
Parkes telescope in NSW, Australia.  The pulsar has been the subject of
regular timing observations at Parkes since 1999 May.  These use the
center beam of the multibeam receiver and the same data-acquisition
system as the survey to collect total-power samples every 0.25\,ms from
96 contiguous frequency channels between 1230 and 1518\,MHz.  The 1-bit
digitized samples are written to tape for subsequent analysis, together
with relevant telescope parameters, including the start time of the
observation referred to the observatory time standard.  The time
on-source is 9\,min on average each observing day.

The timing data are analyzed in standard fashion:  samples from higher
frequency channels are appropriately delayed to compensate for
dispersion in the ionized interstellar medium.  Data from all frequency
channels are then summed and the resulting time series is folded modulo
the predicted topocentric pulsar period to generate an average profile
for each observation.  A sum of several aligned average profiles yields
the ``standard profile'' (Fig.~\ref{fig:prof_res}), with which
individual profiles are correlated to obtain times-of-arrival (TOAs).

The TOAs are analyzed with the {\sc tempo} software
package\footnote{See http://pulsar.princeton.edu/tempo.}: they are
converted to the solar-system barycenter using the JPL~DE200 planetary
ephemeris (\cite{sta90}), and fitted to a Taylor series expansion of
pulse phase by minimizing the weighted rms of timing residuals
(observed minus computed phase).  This yields precise values of pulsar
rotation frequency $\nu$, its derivative $\dot \nu$, and celestial
coordinates.  Many young pulsars, including \psr, experience rotational
instabilities, and in the presence of such ``timing noise'' particular
care must be taken in order to minimize contamination of fitted
parameters.  Following the prescription of Arzoumanian et
al.~(1994)\nocite{antt94} we first ``whiten'' the timing residuals, by
fitting for position, $\nu$, $\dot \nu$, and higher frequency
derivatives as required ($\ddot \nu$ in this case).  The resulting
celestial coordinates and uncertainties are our best estimates of these
parameters and are listed in Table~\ref{tab:parms}, as are $\nu$ and
$\dot \nu$, which we determine from one additional fit with position
held fixed and $\ddot \nu$ set to zero.  We then perform one final fit
which includes $\ddot \nu$.  This is {\em not\/} a term describing the
secular evolution of the neutron star rotation, but results from
rotational instability in the pulsar.  In fact, both the sign and
magnitude of $\ddot \nu$ strongly suggest that it mostly represents
recovery from a glitch that occurred prior to the discovery of the
pulsar (cf. \cite{lyn96}).  The timing residuals corresponding to the
solution presented in Table~\ref{tab:parms}, based on 1.5\,yr of data,
are shown in Figure~\ref{fig:prof_res}.

\medskip
\epsfxsize=8truecm
\epsfbox{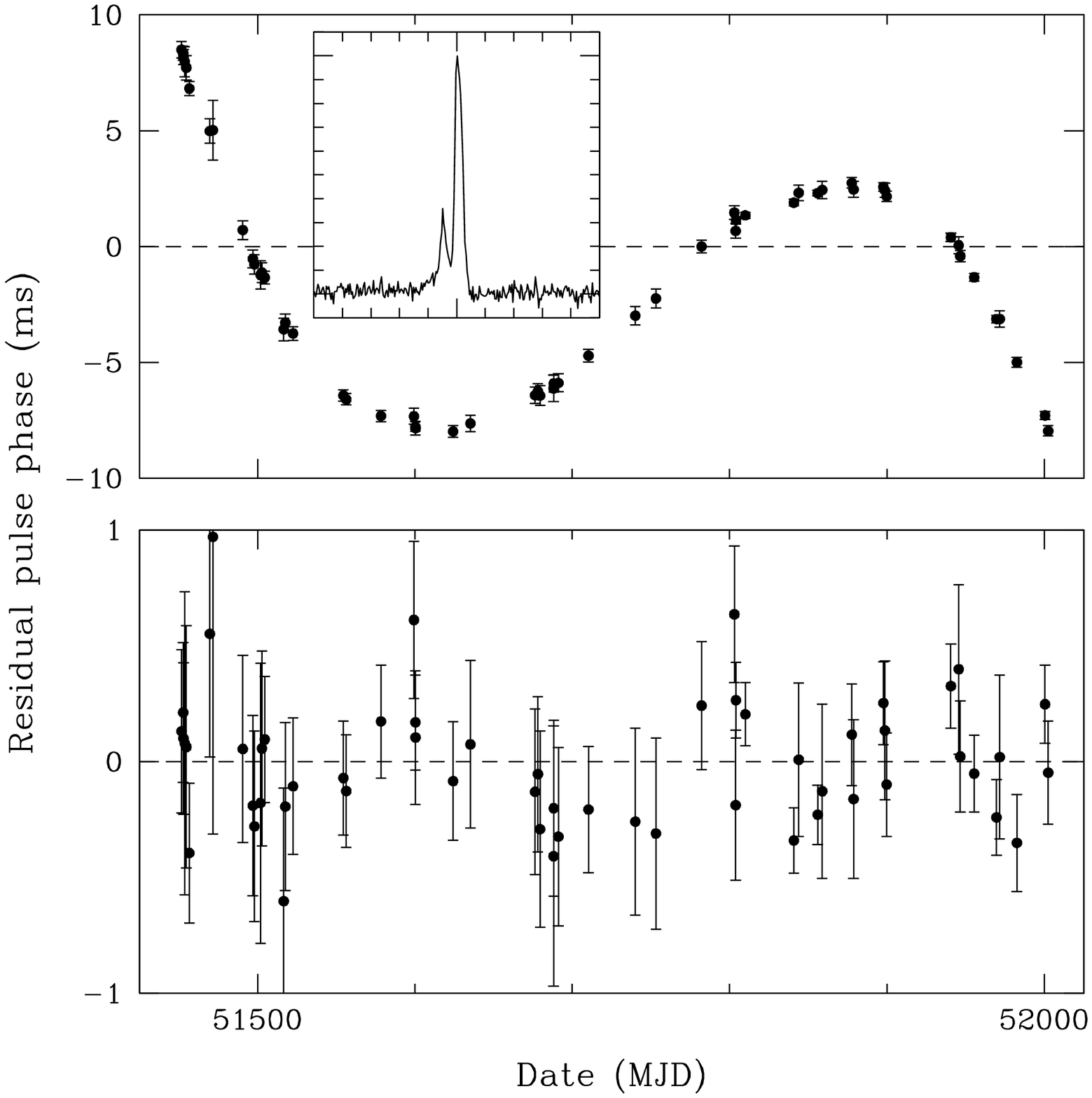}
\figcaption[fig1.eps]{\label{fig:prof_res}
Pulse profile and post-fit timing residuals for \psr.  Corresponding
parameters and weighted rms residuals for ``red'' (top) and ``white''
(bottom) solutions are listed in Table~\ref{tab:parms}.  {\sl Top:
(inset)\/} average pulse profile at 1400\,MHz, with 0.6\,ms
resolution.  The full profile width at half the maximum intensity is
$w_{50} = 3.3$\,ms, and at 10\% it is $w_{10} = 11$\,ms;  {\sl
(main)\/} fit for $\nu$ and $\dot \nu$ only (position held fixed at
best fitted value).  The quasi-cubic trend in residuals indicates
presence of timing noise.  {\sl Bottom:\/} as for upper panel, with
additional fit for $\ddot \nu$ (see \S~\ref{sec:psr}). }
\bigskip

With period $P \equiv 1/\nu =107$\,ms and a high $\dot P$, PSR J1016$-$5857's
inferred characteristic age, $\tau_c = P/2\dot P = 21$\,kyr, is
relatively small and its spin-down luminosity, $\dot E = 4\pi^2 I \dot
P/P^3 = 2.6\times 10^{36}$\,erg\,s$^{-1}$, large.  The surface dipole
magnetic field strength is $B = 3.2\times 10^{19} (P \dot P)^{1/2} =
3.0 \times 10^{12}$\,G (Table~\ref{tab:parms}).  These parameters make
it similar to the small number of ``Vela-like'' pulsars, which have
$\tau_c \sim 10^4$--$10^5$\,yr and $\dot E > 10^{36}$\,erg\,s$^{-1}$.
Prior to the Parkes multibeam survey, only nine such pulsars were
known.  Thus far we have already doubled this number\footnote{See
http://www.atnf.csiro.au/research/pulsar/pmsurv.}.  Pulsars in this
group have high levels of glitch activity, and many are associated with
PWNe and/or SNRs.

\section{Supernova Remnant \snr}\label{sec:snr}

\psr\ is located, at least in projection, near the supernova remnant
\snr\ (MSH~10$-$5{\it 3\/}; \cite{mck+89}; see Fig.~\ref{fig:snr}).
SNR~\snr\ is an incomplete radio shell with non-thermal spectrum and
with significant polarization, interacting with molecular clouds
(\cite{rm86}).  The pulsar is located exactly at the very tip of a
``finger'' of emission on the western edge of the remnant, and $\sim
15'$ from the SNR's approximate geometrical center.

In the Galactic plane, the spatial density of SNRs and pulsars is high,
so the fact that a young pulsar lies near the edge of a SNR does not
necessarily imply that they are associated: the probability of finding
a pulsar lying by chance within $15'$ of a SNR center in this region of
the Galaxy ($270\arcdeg<l<300\arcdeg$) is about 5\% (cf.
\cite{gre00}).  However, the morphology of this particular pairing
appears to bear a resemblance to that of PSR~B1757$-$24 (another
Vela-like pulsar) and its associated SNR~G5.4$-$1.2 (``the Duck'').  In
the latter case, the pulsar has apparently caught up with, penetrated
and overtaken the rim of its SNR, leaving a trail in its wake; the
pulsar is now embedded in a ram-pressure confined (bow-shock) PWN
(\cite{mkj+91}; \cite{fk91}).  Because of the limited resolution and
sensitivity of the MOST data shown in Figure~\ref{fig:snr}, it cannot
yet be determined whether a similar association exists between
\psr\ and SNR~\snr.  Nevertheless, the location of \psr\ at the tip of a
bright structure connected to \snr\ is suggestive of a physical
association between the pulsar and SNR.

\epsfxsize=8truecm
\epsfbox{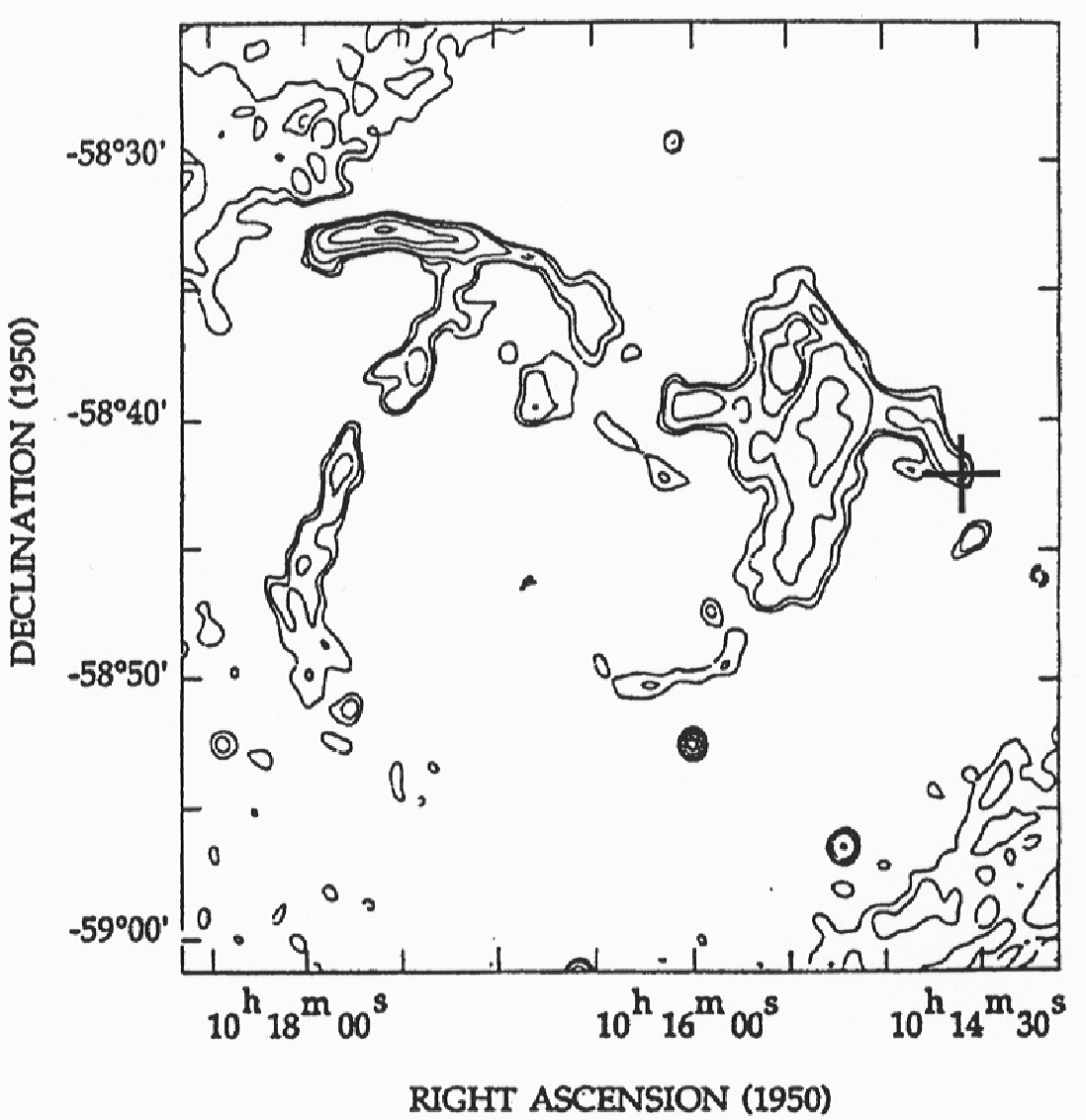}
\figcaption[fig2.eps]{\label{fig:snr}
MOST radio map at 843\,MHz of SNR~\snr\ (\cite{mck+89}), with a spatial
resolution of $45''$.  The position of \psr, known with sub-arcsecond
accuracy (Table~\ref{tab:parms}), is indicated with a large cross at
the tip of the protuberance on the western edge of the SNR. }
\bigskip

If the pulsar and SNR are associated, the pulsar is at the distance of
the SNR, $d=3$\,kpc, on the near side of the Carina spiral arm, as
determined from optical and CO line observations with an uncertainty of
only $\approx 20\%$ (\cite{rm86}).  This distance is smaller than that
inferred from the pulsar's dispersion measure (DM) and a model for the
free electron distribution in the Galaxy, $d = 9_{-2}^{+3}$\,kpc
(\cite{tc93}).  However, the model is at its most uncertain along
spiral arm tangents, the direction of \psr, so that the discrepancy may
not be as serious as it appears.  A similar example is provided by the
recently discovered $\sim 1600$-yr-old pulsar J1119$-$6127
(\cite{ckl+00}) and its clearly associated SNR G292.2$-$0.5
(\cite{cgk+01}; \cite{pkc+01}), where the model puts the pulsar at
$d>30$\,kpc, when in fact it lies at $d\la8$\,kpc.  Hypothesizing the
\psr/SNR~\snr\ association, we consider $d=3$\,kpc hereafter.

The age of \snr\ is estimated by Ruiz \& May (1986)\nocite{rm86} as
$\sim 10$\,kyr, by fitting line profiles observed from an optical
filament to models of interstellar radiative shocks.  This is similar
to the pulsar's characteristic age, $\tau_c = 21$\,kyr.  However,
pulsar characteristic ages are not necessarily accurate estimates of
true age $\tau$: e.g., in the B1757$-$24/G5.4$-$1.2 system, $\tau \gg
\tau_c$ (\cite{gf00}), while for the pulsar in SNR~G11.2$-$0.3, $\tau
\ll \tau_c$ (\cite{krv+01}).  Still, if the \psr/\snr\ association is
real, and the age of the system is $\sim \tau_c$, we can estimate the
velocity required for the pulsar to move ballistically from its
putative birthplace (SNR center) to its present location: $v \sim 500
(d/3\,\mbox{kpc})(21\,\mbox{kyr}/\tau)$\,km\,s$^{-1}$.  This velocity
is within the range expected for young neutron stars (\cite{ll94}), but
is highly supersonic in the local ISM.  A characteristic bow-shock PWN
is then expected, as is observed for PSR~B1757$-$24 (\cite{gf00}).
Such a bow-shock nebula, of typical extent $\la 1'$, could not be seen
in Figure~\ref{fig:snr}.  Furthermore, the larger-scale finger
connecting to the SNR need not point in the direction of motion, since
the ``wake'' left behind by the pulsar can be distorted by ISM motions
and turbulence.  Finally, in comparing the \psr/\snr\ and
PSR~B1757$-$24/G5.4$-$1.2 systems, we note that the pulsars are
uncannily alike:  $P=107/128$\,ms, $\tau_c=21/16$\,kyr, $\dot E =
2.6/2.5 \times 10^{36}$\,erg\,s$^{-1}$, $B=3/4 \times 10^{12}$\,G.

\section{X-ray Observations}\label{sec:einstein}

Energetic young pulsars such as \psr\ often produce detectable
high-energy emission, especially if their distance is not too large (as
we have argued in the previous section).  A search of the HEASARC
archive for serendipitous high-energy observations of \psr\ has yielded
one pointing at SNR~\snr, obtained with the Imaging Proportional
Counter (IPC) on-board the {\sl Einstein Observatory}. The IPC is
sensitive to 0.1--4.5\,keV X-rays over its $1\arcdeg \times 1\arcdeg$
field-of-view with an image resolution of $1'$--$3'$. The standard
processed image from this 1.9\,ks exposure (Seq. I5288), acquired on
1980 February 5, is shown in Figure~\ref{fig:einstein}.  The pulsar
position is near the center of the IPC field and coincides with an
excess of localized X-ray emission, the most significant in the field.
This emission is consistent, given the limited number of counts, with a
point source, with possible extended emission.  From the original event
data files, we extracted 134 counts from a $r= 3\farcm0$ circle around
the peak of emission and estimated the background counts within a
surrounding annulus at $3\farcm2 < r < 6\farcm2$.  The 45
background-subtracted counts yield a source significance of
4.8\,$\sigma$ and a centroid which is $\sim 1\farcm8$ offset from the
radio position.  While much larger than the nominal \einstein\ pointing
error ($\la 10''$), this offset is consistent with the IPC instrumental
measurement error and the small number statistics (e.g., \cite{har+90}).

To estimate the luminosity of the source we assume a typical pulsar
power-law model with photon index $\approx 2$, as might be appropriate
for magnetospheric emission or origin in a non-thermal nebula. A
measure of the Galactic neutral hydrogen column density to the pulsar
is estimated from the H{\sc i} study of Dickey \&
Lockman~(1990)\nocite{dl90}\footnote{Accessible via the {\sl nH\/} tool
at http://heasarc.gsfc.nasa.gov/docs/corp/tools.html.}.  This gives
$N_H \approx 10^{22}$\,cm$^{-2}$, consistent with the rule of thumb of
0.1 free electrons per H atom and the pulsar's DM.  For isotropic
emission, we obtain an unabsorbed X-ray luminosity of L$_{x,\rm
0.1-4.5\,kev} \sim 10^{33} (d/3\,\mbox{kpc})^2$\,erg\,s$^{-1}$.  This
represents an efficiency for conversion of rotation-driven luminosity
into X-rays of $L_x/\dot E \sim 5\times 10^{-4}$ in the
\einstein\ band.  Such an efficiency for the point source and any
compact nebula surrounding it is consistent within uncertainties with
that observed for young pulsars (e.g., \cite{sw88}).

\epsfxsize=8truecm
\epsfbox{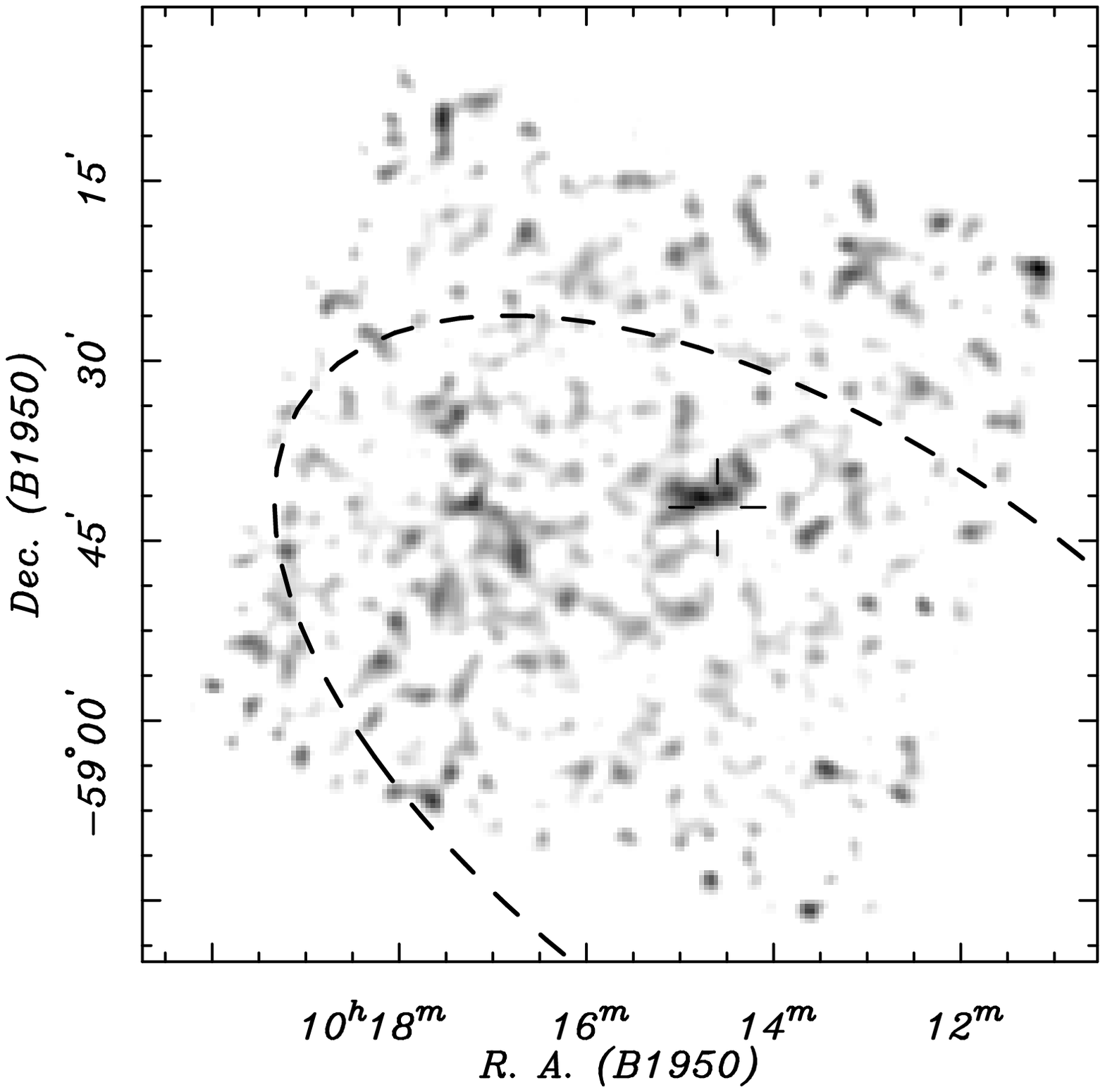}
\figcaption[fig3.eps]{\label{fig:einstein}
\einstein\ IPC observation of the \psr/SNR \snr\ field: a possibly
extended source is detected near the position of the pulsar
(cross-hairs; see \S~\ref{sec:einstein}). The ellipse overlaid
represents the approximate 2\,$\sigma$ contour of positional
uncertainty for the \egret\ source \egr\ (\cite{hbb+99}; see
\S~\ref{sec:egret}), which has an effective 2\,$\sigma$ uncertainty
radius $43'$. }
\bigskip

Given the limitations of the detection, the origin of the observed
emission is not certain at present.  Possible contributions include
emission from: a point source (e.g., pulsed magnetospheric emission); a
compact nebula (bow-shock or static); the SNR; or even interaction of
the SNR with a molecular cloud (cf. \cite{rm86})\footnote{Only one
cataloged object exists within $2'$ of the pulsar position, the A2 star
HD~302636, $1'$ away.  If this star were a supergiant it would lie
outside the Galaxy, while if it is a main-sequence A2V star it lies at
$\la 1$\,kpc.  At such a distance, the flux measured from the
\einstein\ source would imply a luminosity in excess of that known for
any A2V star.}.  However, the inferred flux is in excess of the thermal
emission expected from a $\sim 20$\,kyr-old neutron star for a
reasonable $N_H$, while the measured flux is approximately as expected
from a PWN in such a pulsar system.  Also, if the extended emission is
real, it appears to point back towards the center of the SNR,
suggestive of a ram-pressure confined wind nebula.

\section{$\gamma$-ray Source \egr}\label{sec:egret}

As shown in Figure~\ref{fig:einstein}, \psr\ and the \einstein\ X-ray
source are within the $\approx 43'$ error box of \egret\ source \egr.
This $\gamma$-ray source has flux $(3.3\pm0.6)\times 10^{-7}$
photons\,cm$^{-2}$\,s$^{-1}$ ($>100$\,MeV), and photon index $\Gamma =
2.32\pm0.13$ (\cite{hbb+99}).  This photon index is slightly larger
than for most $\gamma$-ray pulsars, but consistent with those of the
Crab ($\Gamma = 2.19$), and the Vela-like PSR~J2229+6114 ($\Gamma =
2.24$), which is almost certainly the source of the $\gamma$-rays from
3EG~J2227+6122 (\cite{hcg+01}).  \egr\ is possibly a composite source
(\cite{hbb+99}), in which case the photon index of any pulsar component
may be different from the nominal value, and its flux will be lower
than that listed.  The source is not obviously variable
(\cite{mmct96}), and the flux integrated from 100\,MeV to 10\,GeV with
the nominal photon index corresponds to an isotropic luminosity
$L_\gamma \sim 1.6\times 10^{35} (d/3\,\mbox{kpc})^2$\,erg\,s$^{-1}$,
or $\sim 0.06 \dot E$.

The chance probability for superposition of a Vela-like pulsar and an
\egret\ error box in the Galactic plane is not negligible: in the 1500
square degrees covered by the Parkes multibeam survey, where there are
about 10 Vela-like pulsars known, \egret\ error boxes cover 24 square
degrees or 1.6\% (\cite{hbb+99}).  The chance probability that at least
one of those pulsars is aligned with one of the error boxes is
therefore $1-(1-0.016)^{10}=0.15$.  The actual probability is somewhat
smaller if one considers only non-variable \egret\ sources, but it is
nevertheless significant.

However, both $L_\gamma/\dot E \sim 0.06$ and $L_\gamma/L_x \sim 100$
are approximately as expected for a pulsar of the age and $\dot E$ of
\psr\ (e.g., \cite{tbb+99}, and \cite{sw88}), as is the lack of
variability in $\gamma$-rays.  Also, every pulsar with a value of $\dot
E/d^2$ greater than that of \psr\ has been detected in $\gamma$-rays
(although the distances of \psr\ as well as those of other pulsars are
substantially uncertain).  Altogether, this makes an association
between \psr\ and \egr\ possible and intriguing.  The proof that
\egr\ is the $\gamma$-ray counterpart of \psr\ would come from the
detection of $\gamma$-ray pulsations.  This will be difficult with
\egret\ data, since the timing noise of the pulsar and the likelihood
of glitches (\S~\ref{sec:psr}) prevent the reliable backwards
extrapolation of the ephemeris required to search for $\gamma$-ray
pulsations.

\section{Discussion and Future Work}\label{sec:disc}

We have discovered an energetic and youthful pulsar, apparently located
just outside SNR~\snr.  Based on its spin characteristics, \psr\
belongs to the rare class of Vela-like pulsars, those with $\tau_c \sim
10^4$--$10^5$\,yr and $\dot E > 10^{36}$\,erg\,s$^{-1}$.  These pulsars
are of particular interest, as many are embedded in PWNe and are
observed to be interacting with their SNRs.  Additionally, such pulsars
are very efficient at converting their spin-down luminosity into
$\gamma$-rays, and form the majority of \egret\ sources identified with
pulsars.

We have argued, based on its positional coincidence with one notable
feature in the SNR, that \psr\ may be interacting with \snr, and
therefore that the two objects are physically associated.  Further
supporting this notion is the detection of an X-ray source nearby
\psr\ and the positionally coincident unidentified $\gamma$-ray source
\egr.  The luminosities of both X- and $\gamma$-ray sources, if located
at the distance of \snr, are consistent with provenance from \psr.

Confirming this picture will require high-resolution radio and X-ray
observations that show the proposed interaction between pulsar and
SNR.  If a bow-shock is detected, its detailed study may constrain the
pulsar velocity, age, spin-evolution, and ambient medium (cf.
\cite{gf00}; \cite{ocw+01}).  Confirmation of the nature of the
\egret\ source may have to await future missions such as {\sl
GLAST\/}.

\acknowledgments

The Parkes radio telescope is part of the Australia Telescope which is
funded by the Commonwealth of Australia for operation as a National
Facility managed by CSIRO.  This research has made use of data obtained
from the High Energy Astrophysics Science Archive Research Center
(HEASARC), provided by NASA's Goddard Space Flight Center, and of the
SIMBAD database, operated at CDS, Strasbourg, France.  FC acknowledges
support from NASA grants NAG~5-9095 and ADP NAG~5-9120.  IHS is a
Jansky Fellow.  EVG is supported by NASA LTSA grant NAG5-22250.


\clearpage

\begin{deluxetable}{ll}
\footnotesize
\tablecaption{\label{tab:parms}Parameters of \psr\ }
\tablecolumns{2}
\tablewidth{0pc}
\tablehead{
\colhead{Parameter}    &
\colhead{Value}     \nl}
\startdata
R.A. (J2000)\dotfill                     & $10^{\rm h}16^{\rm m}21\fs16(1)$  \nl
Decl. (J2000)\dotfill                    & $-58\arcdeg57'12\farcs1(1)$       \nl
Rotation frequency, $\nu$ (s$^{-1}$)\dotfill        & 9.31274229245(4)       \nl
Period, $P$ (s)\dotfill                  & 0.1073797565311(5)                \nl
Frequency derivative, $\dot \nu$ (s$^{-2}$)\dotfill & 
                                               $-6.991787(7)\times 10^{-12}$ \nl
Period derivative, $\dot P$\dotfill      & $8.061818(8)\times 10^{-14}$      \nl
Second frequency derivative, $\ddot \nu$ (s$^{-3}$)\tablenotemark{a}\dotfill & 
                                  $1.03(2)\times 10^{-22}$                   \nl
Epoch (MJD)\dotfill                            & 51730.0                     \nl
Dispersion measure, DM (cm$^{-3}$\,pc)\dotfill & 394.2(2)                    \nl
Data span (MJD)\dotfill                        & 51451--52002                \nl
Rms timing residual (red/white) (ms)\tablenotemark{b} \dotfill  & 4.3/0.24   \nl
Flux density at 1400\,MHz (mJy)\dotfill        & 0.46(5)                     \nl
Derived parameters:                            &                             \nl
~~~Galactic longitude, $l$ ($\deg$)\dotfill    & $284.08$                    \nl
~~~Galactic latitude, $b$ ($\deg$)\dotfill     & $-1.88$                     \nl
~~~Distance, $d$ (kpc)\tablenotemark{c}\dotfill& 3?                          \nl
~~~Characteristic age, $\tau_c$ (yr)\dotfill   & $2.1\times10^4$             \nl
~~~Spin-down luminosity, $\dot E$ (erg\,s$^{-1}$)\dotfill 
                                                      & $2.6 \times 10^{36}$ \nl
~~~Magnetic field strength, $B$ (G)\dotfill    & $3.0 \times 10^{12}$        \nl
\enddata
\tablecomments{Numbers in parentheses represent 1\,$\sigma$
uncertainties in the least-significant digits quoted.}
\tablenotetext{a}{fitted for separately (not a secular term) --- see
\S~\ref{sec:psr}.}
\tablenotetext{b}{see Figure~\ref{fig:prof_res}.}
\tablenotetext{c}{see \S~\ref{sec:snr}.}
\end{deluxetable}

\end{document}